\documentclass[fleqn,usenatbib]{mnras}

\usepackage{newtxtext,newtxmath}

\usepackage[T1]{fontenc}

\usepackage{url}
\usepackage{subfigure}
\usepackage{longtable}
\usepackage{lscape}
\usepackage{multirow}
\usepackage{color}
\usepackage{textcomp}
\usepackage{natbib}
\usepackage{footnote}
\usepackage{verbatim}

\setcitestyle{notesep={;}}

\DeclareRobustCommand{\VAN}[3]{#2}
\let\VANthebibliography\thebibliography
\def\thebibliography{\DeclareRobustCommand{\VAN}[3]{##3}\VANthebibliography}

\usepackage{graphicx}	
\usepackage{amsmath}	

\usepackage{amssymb}

\title[ molecular clouds catalogue]{A large catalogue of molecular clouds in the southern sky}

\author[Guo et al.]{
H.-L. Guo, $^{1}$ B.-Q. Chen,$^{1}$\thanks{E-mail: bchen$@$ynu.edu.cn (BQC)}  
and X.-W. Liu$^{1}$
\\
$^{1}$South-Western Institute for Astronomy Research, Yunnan University, Chenggong District, Kunming 650091, P.\,R. China\\
}

\date{Accepted XXX. Received YYY; in original form ZZZ}
\pubyear{2021}
\makeatother

\begin{document}
\label{firstpage}
\pagerange{\pageref{firstpage}--\pageref{lastpage}}

\maketitle
\begin{abstract}
We present a large catalogue of molecular clouds with accurate distance estimates in the southern sky. Based on the three-dimensional (3D) dust extinction map and the best-fit extinction and distance information of over 17 million stars presented in Guo et al., we have identified 250 dust/molecular clouds in the southern sky using a hierarchical structure identification algorithm. Amongst them, 71 clouds locating at high Galactic latitudes ($b < -10$\degr). We have estimated the distances to the clouds by fitting the extinction versus distance profiles of the lines of sight overlapping with the clouds using a simple Gaussian dust distribution model. The typical uncertainties of the distances are less than 7 per\,cent. We also provide the physical properties  of the individual clouds, including the linear radius, mass and surface mass density.

\end{abstract}


\section{Introduction}
Molecular clouds play a crucial role in the processes of star formation and galaxy evolution \citep{Blitz1999}. A large catalogue of molecular clouds with robust property measurements is fundamental for the studies of the physical nature and evolution of molecular clouds themselves and understanding the process of star formation and ultimately of galaxy evolution \citep[e.g.][]{Larson1981, Lombardi2010, Wang2015, Li2016, Rice2016, Lada2020}.

Many efforts have been made to identify molecular clouds in the Milky Way and obtain their physical parameters. Most of those works are based on the radio observations of the CO emission. Based on the CO data obtained with the 5\,m telescope of the Millimeter Wave Observatory, \citet{Magnani1985} identified 57 clouds at high Galactic latitudes and provided properties such as distance, size, mass, density and extinction of some of the clouds. Based on the third Galactic quadrant CO survey obtained with the 1.2\,m Southern Millimeter-wave Telescope  \citep{May1993}, \citet{May1997} identified 177 molecular clouds in the southern outer Galaxy and measured their distances, radii and masses. They found power-law relations between the cloud line widths and sizes, as well as between the densities and sizes. \citet{Dame2001} combined the available large-scale CO surveys into a composite survey that covers most of the sky with significant CO emission. Based on the data of Dame et al., \citet{Rice2016} and \citet{Miville2017} identified and measured the basic properties of molecular clouds across the Milky Way. \citet{Rice2016} identified 1,064 clouds in the Galactic plane using a dendrogram-based decomposition algorithm. \citet{Miville2017} recovered 8,107 molecular clouds in the entire Galactic plane by Gaussian decomposition followed by a hierarchical cluster identification method.

Dust observation serves as a complementary approach to trace the molecular gas and probe the properties of molecular clouds \citep{Wolf1923, Lada1994}. Based on the Digitized Sky Survey \uppercase\expandafter{\romannumeral1} \citep{Lasker1994}, \citet{Dobashi2005} produced a two-dimensional (2D) $A_{V}$ extinction  map of the Galactic disk and identified 2,448 dark clouds from the map. The catalogue was later improved by \citet{Dobashi2011}, who identified 7,614 clouds across the sky based on a 2D extinction map derived from  the data of the 2 Micron All Sky Survey \citep[2MASS;][]{Skrutskie2006}. Based on the 2MASS data and the Besan\c{c}on Galactic model \citep{Robin2003}, \citet{Marshall2009} mapped the three-dimensional (3D) extinction distributions toward over 1,500 dark clouds and calculated the distances and masses for 1,259 of them. Based on the 2MASS data, \citet{Lada2010}, \citet{Lombardi2010} and \citet{Lombardi2011} presented near-infrared (IR) extinction maps toward nearby molecular clouds and estimated distances and masses of them. \citet{Schlafly2014} determined the distances of clouds identified by \citet{Magnani1985} and of some other well-studied nearby clouds by the method of 3D extinction mapping based on the data of PanSTARRS-1 \citep[][ PS1]{Chambers2016}. Using similar techniques, \citet{Zucker2019, Zucker2020} measured the distances of local molecular clouds by combining the
optical and near-IR photometries with the Gaia Data Release
2 \citep[Gaia DR2;][]{Gaia2018}. Recently, \citet{Chen2020} presented a large catalogue of 567 clouds in the Galactic plane identified from the 3D extinction maps of \citet{Chen2019} and provided their accurate distances and other physical parameters.

Most of the aforementioned studies dealt with molecular clouds in the Galactic disk of low Galactic latitudes and of regions of high Galactic latitudes in the northern sky. Only few works have analysed the molecular clouds in the southern sky, especially of regions of $b < -10\degr$\ . \citet{Guo2021a} have presented 3D extinction maps of the southern sky based on the multiband photometry of the SkyMapper Southern Survey Data Release 1 \citep[SMSS DR1;][]{Wolf2018}, Gaia DR2, 2MASS, and the Wide-Field Infrared Survey Explorer \citep[WISE;][]{Wright2010}, and the stellar distances calculated from the Gaia DR2 parallaxes \citep{Bailer2018}. The work of Guo et al. provides us with an opportunity to fill the gap.

In the current work, we identify molecular clouds in the southern sky and calculate their accurate distances and other physical properties based on the data of \citet{Guo2021a}. The structure of the paper is as following. The data and method are described in Section 2 and 3, respectively. We present our results in Section 4 and discuss them in Section 5. A summary is given in
Section 6.

\section{Data}
\label{data}
The basic data adopted in the current work are from \citet{Guo2021a}, including the 3D dust extinction maps of the southern sky and a catalogue containing distances and extinction values of over 17 million stars. The data can be accessed at {\url{https://registry.china-vo.org/resource/101032}}.

The extinction maps of Guo et al. have a sky coverage of $\sim$ 14,000 deg$^2$ within the footprint of the SMSS DR1 \citep{Wolf2018}. \citet{Guo2021a} adopted only stars that are detected in all the SkyMapper $g$, $r$ and $i$ bands with photometric uncertainties smaller than 0.08\,mag in all the three bands. Their maps cover the
entire southern sky (Declination $\delta < 0\degr$) except for a few regions close to the Galactic plane and some isolated fields due to the missing of the SMSS DR1 data. The resolution of the maps is 6.9 arcmin at low Galactic latitudes ($|b| < 10\degr$) and increases to 27.5 arcmin at $b < -30\degr$.\ The maps have a typical distance limit of $\sim$ 5\,kpc and a distance resolution of 0.2\,kpc.

The extinction values of the individual stars were estimated by the best-fit SED algorithm using the multi-band photometric data from the optical to the near-IR bands. The typical uncertainties are about 0.07\,mag in $E(B - V)$ \citep{Guo2021a}. The distances of stars were adopted from the work of \citet{Bailer2018}, who provide distances of 1.3 billion stars based on the Gaia DR2 parallax measurements using a simple Bayesian approach. \citet{Guo2021a} excluded stars with Gaia DR2 parallax uncertainties larger than 20 per\,cent, leading to a sample of over 17 million stars with robust distance and extinction values.

\begin{figure*}
\centering

\includegraphics[width=1.0\textwidth, angle=0]{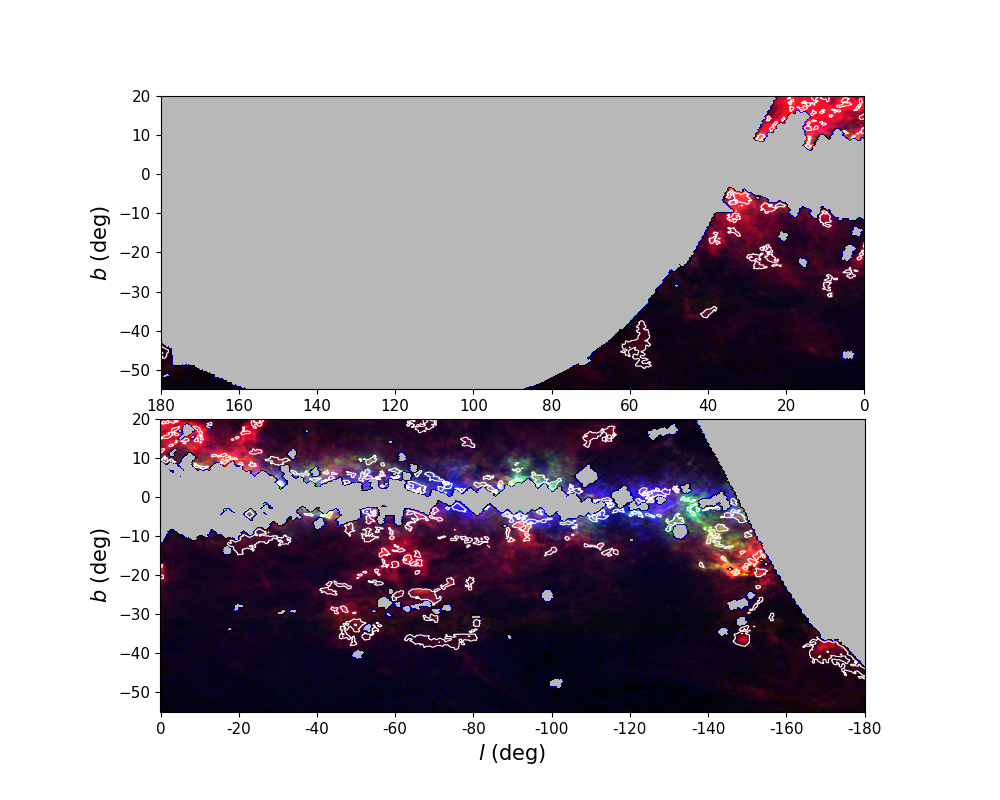}
\vspace{-1.cm}

\caption{Spatial distributions of the molecular clouds identified in this work. The boundaries of the individual molecular clouds are marked by white polygons. The 3D dust extinction maps of \citet{Guo2021a} are represented by the background three-colour composite images, in which the red, green and blue scales correspond to the extinction values in the distance slices 0--800, 800--1600, and 1600--5000\,pc from the Sun, respectively. The grey pixels, which are mostly located at the Galactic plane, are regions not covered by the extinction maps of \citet{Guo2021a}.}
\label{contour}
\end{figure*}

\begin{table*}
  \centering
  \caption{Catalogue of molecular clouds.}
  \begin{tabular}{rrrr  *{6}{r} }
    \hline
    \hline
    ID  & $l$        & $b$       & $\Omega$  & $r$  & $d_0$            & $\delta d$      & $\delta{A_r}$  & $M$           & $\Sigma$               \\
        & ($\degr$)  & ($\degr$) & (deg$^2$) & (pc) & (pc)             & (pc)            & (mag)          & ($M_{\odot}$) & $ (M_{\odot}$\,pc$^{-2}$) \\
    \hline
1 & 57.63 & $-$43.12 & 28.115 & 9.3 & 179 $\pm$ 4 & 70 $\pm$ 4 & 0.22 & 1494.0 & 5.4 \\
2 & 188.65 & $-$39.66 & 50.676 & 16.8 & 239 $\pm$ 4 & 126 $\pm$ 2 & 0.29 & 13024.7 & 14.8 \\
3 & 287.4 & $-$36.6 & 31.583 & 11.1 & 200 $\pm$ 5 & 69 $\pm$ 6 & 0.22 & 1583.2 & 4.1 \\
4 & 211.1 & $-$35.88 & 13.753 & 5.2 & 143 $\pm$ 6 & 41 $\pm$ 10 & 0.32 & 3401.1 & 39.7 \\
5 & 310.87 & $-$34.17 & 18.114 & 8.5 & 203 $\pm$ 4 & 53 $\pm$ 9 & 0.38 & 3112.7 & 13.7 \\
6 & 40.29 & $-$35.47 & 4.71 & 9.5 & 445 $\pm$ 12 & 171 $\pm$ 14 & 0.36 & 2024.1 & 7.1 \\
7 & 305.59 & $-$34.33 & 7.654 & 5.0 & 185 $\pm$ 8 & 45 $\pm$ 10 & 0.31 & 1259.1 & 15.8 \\
8 & 278.87 & $-$32.58 & 3.35 & 3.7 & 208 $\pm$ 40 & 47 $\pm$ 21 & 0.21 & 218.9 & 5.0 \\
9 & 290.32 & $-$30.98 & 3.531 & 3.4 & 184 $\pm$ 15 & 55 $\pm$ 14 & 0.29 & 590.7 & 16.2 \\
10 & 9.18 & $-$28.4 & 8.923 & 5.0 & 170 $\pm$ 6 & 51 $\pm$ 8 & 0.39 & 1188.2 & 15.1 \\

    ... & ...        & ...       & ...       & ...  & ...              & ...             & ...            & ...           & ...                    \\
241 & 260.52 & $-$5.66 & 0.498 & 13.5 & 1943 $\pm$ 68 & 242 $\pm$ 85 & 1.05 & 9346.4 & 16.3 \\
242 & 276.7 & $-$3.09 & 0.18 & 9.2 & 2199 $\pm$ 242 & 389 $\pm$ 348 & 1.08 & 4595.8 & 17.4 \\
243 & 299.81 & 5.01 & 0.996 & 24.5 & 2490 $\pm$ 295 & 940 $\pm$ 2301 & 1.7 & 27417.5 & 14.6 \\
244 & 237.23 & $-$4.16 & 0.977 & 24.1 & 2476 $\pm$ 239 & 825 $\pm$ 734 & 1.29 & 23392.3 & 12.8 \\
245 & 296.01 & $-$3.98 & 1.776 & 36.4 & 2776 $\pm$ 177 & 605 $\pm$ 393 & 0.69 & 53535.6 & 12.8 \\
246 & 276.12 & $-$2.57 & 0.14 & 9.2 & 2492 $\pm$ 264 & 452 $\pm$ 366 & 1.33 & 4218.3 & 15.9 \\
247 & 282.97 & 1.62 & 0.9 & 27.2 & 2913 $\pm$ 307 & 775 $\pm$ 417 & 1.6 & 44903.1 & 19.3 \\
248 & 235.37 & $-$4.58 & 0.718 & 21.1 & 2529 $\pm$ 191 & 492 $\pm$ 356 & 1.0 & 17269.9 & 12.4 \\
249 & 286.92 & 3.34 & 1.577 & 35.1 & 2839 $\pm$ 232 & 525 $\pm$ 2944 & 0.6 & 32475.8 & 8.4 \\
250 & 254.12 & 1.69 & 4.937 & 77.0 & 3520 $\pm$ 133 & 628 $\pm$ 269 & 0.6 & 187425.6 & 10.1 \\

    \hline
  \end{tabular}
  \parbox{\textwidth}{\footnotesize \baselineskip 3.8mm
    {\it Note.} The Table is available in its entirety in the online version of this paper and also from the link {\url{http://paperdata.china-vo.org/guo/cloud/cloud_para.fits}}}
\label{catalogue}
\end{table*}

\begin{figure*}
\centering
\includegraphics[width=0.48\textwidth, angle=0]{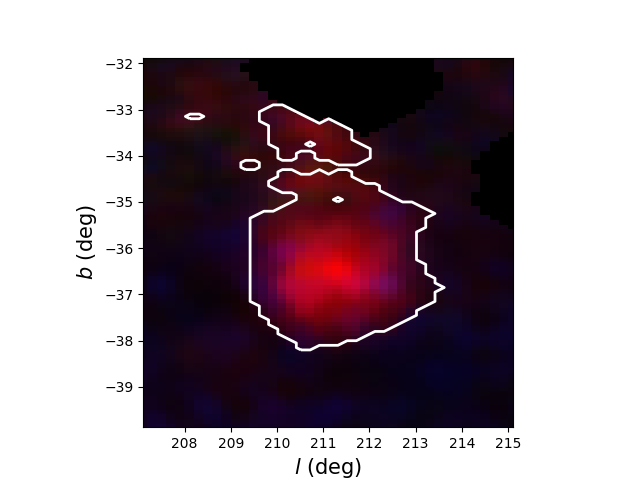}
\includegraphics[width=0.48\textwidth, angle=0]{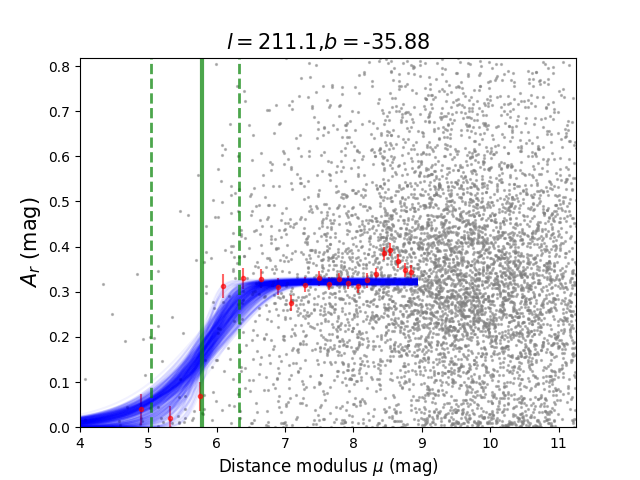}
\includegraphics[width=0.48\textwidth, angle=0]{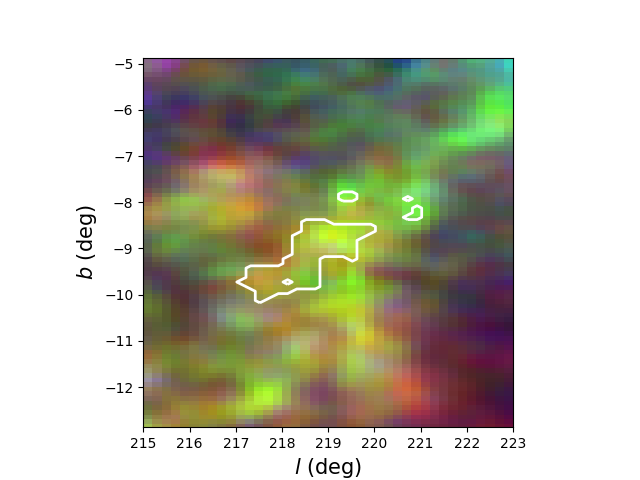}
\includegraphics[width=0.48\textwidth, angle=0]{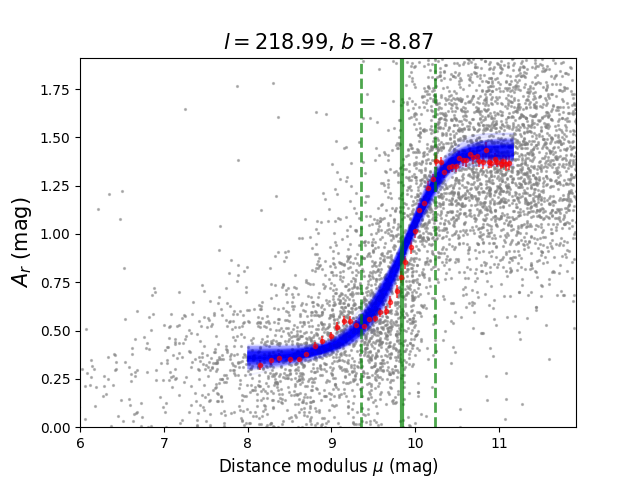}

\caption{Distance determinations for two example molecular clouds: Cloud No.\,4 (upper panel) and No.\,180 (Mon\,R2; bottom panel). In the left panel, we show the sky coverage of the clouds by the white polygons. The 3D dust extinction maps of \citet{Guo2021a} are illustrated by the background three-colour composite images, in which the red, green and blue scales correspond to the extinction values of  $A_r$ in the distance slices 0--800, 800--1600, and 1600--5000\,pc from the Sun, respectively. In the right panel, we show the $r$-band extinction profiles of the sightlines overlapping with the example clouds. The red circles and errorbars are respectively the median values and standard errors in the individual distance bins of bin-size 50\,pc. Blue curves are the best-fit extinction profiles $A_r (d)$ of the 300 randomly generated star samples. Vertical solid and dashed lines mark the resultant distance $d_0$ and width $\delta d$ of the clouds, respectively.}
\label{distance}
\end{figure*}

\begin{figure*}
\centering
\includegraphics[width=0.48\textwidth, angle=0]{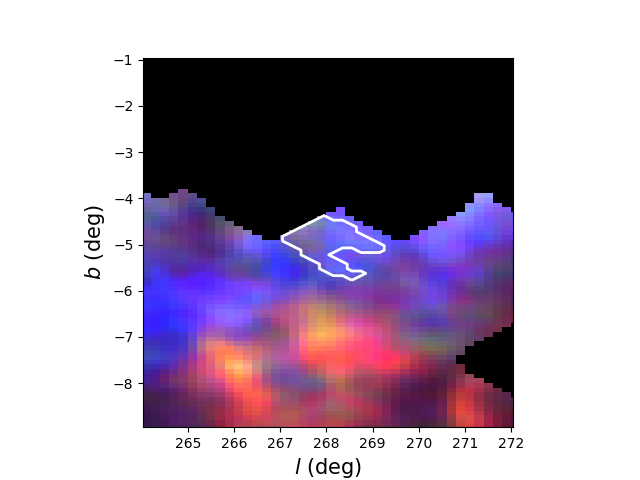}
\includegraphics[width=0.48\textwidth, angle=0]{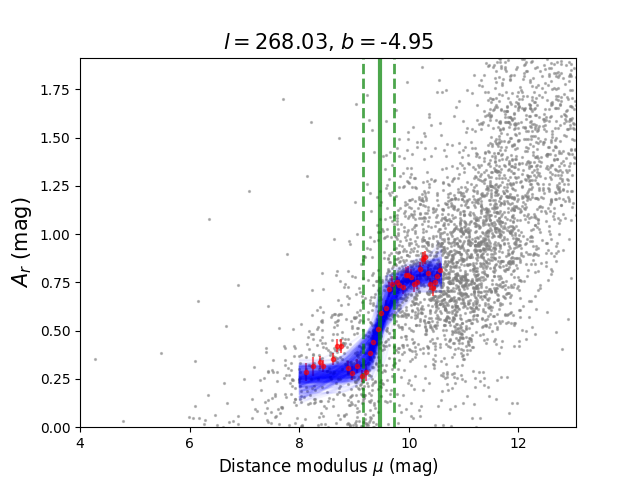}
\includegraphics[width=0.48\textwidth, angle=0]{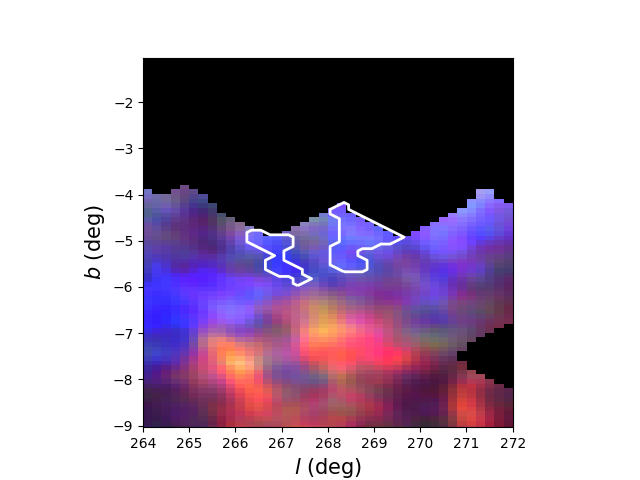}
\includegraphics[width=0.48\textwidth, angle=0]{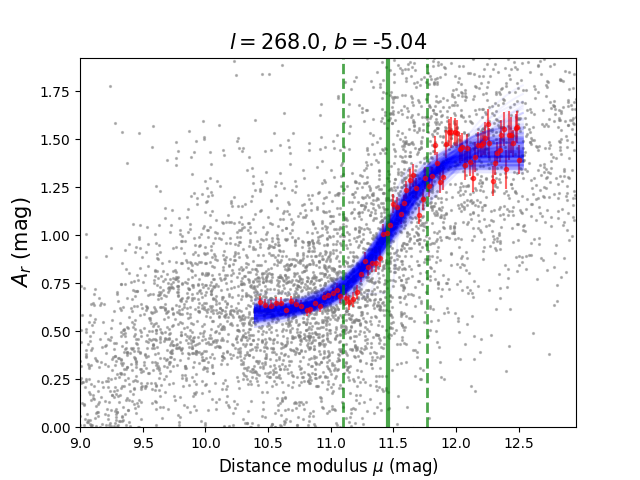}
\caption{Same as Fig.\,\ref{distance} but for Clouds No.\,182 (upper panel) and 237 (bottom panel), that are only partly covered by the data of \citet{Guo2021a}.}
\label{twojump}
\end{figure*}

\begin{figure*}
\centering
\includegraphics[width=0.48\textwidth, angle=0]{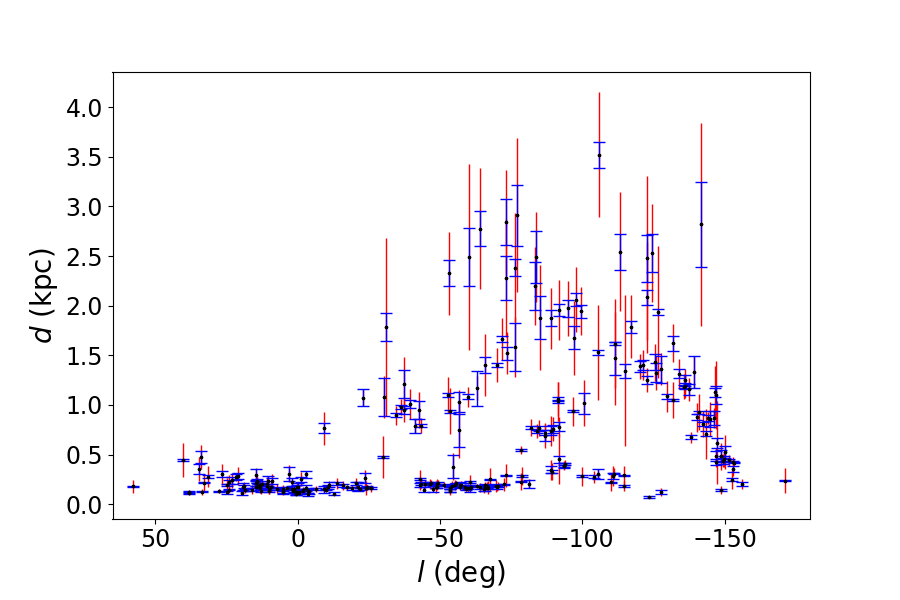}
\includegraphics[width=0.48\textwidth, angle=0]{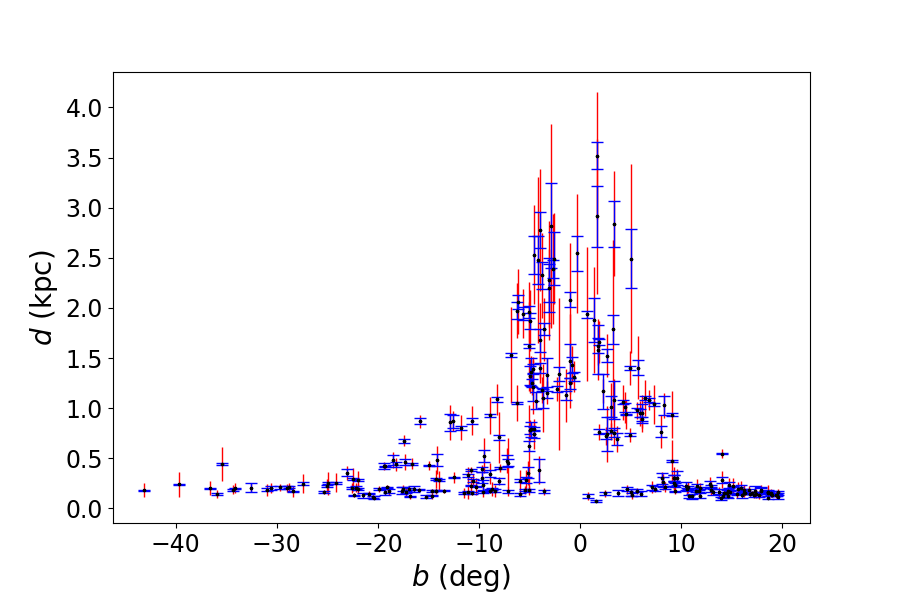}

\caption{Distances of clouds identified in this work plotted against their Galactic longitudes (left panel) and Galactic latitudes (right panel). The blue and red error bars represent the distance uncertainties and the derived widths/depths ($\delta d$) of the clouds, respectively.}
\label{lb_dis}
\end{figure*}

\begin{figure}
\centering
\includegraphics[width=0.45\textwidth, angle=0]{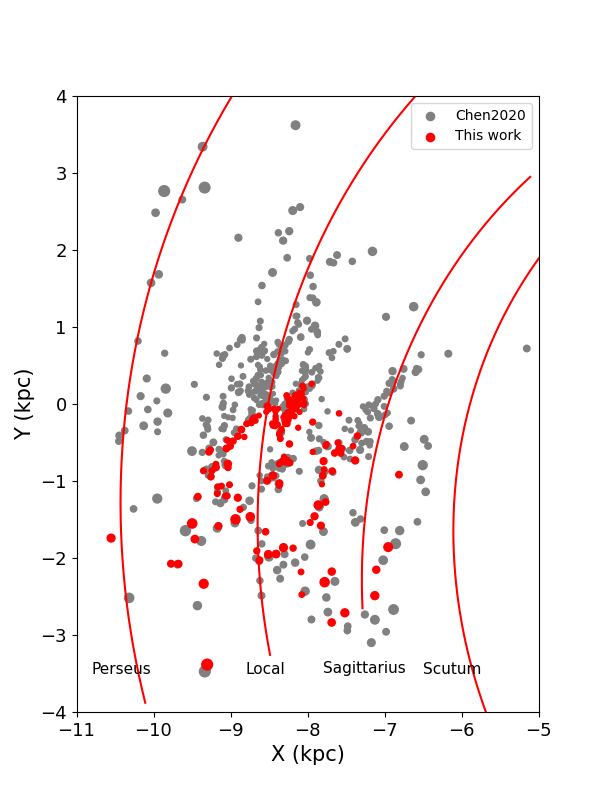}
\caption{Spatial distribution of the molecular clouds identified in the
current work (red dots) and \citet[][ grey]{Chen2020} in the Galactic
coordinates. The sizes of the dots indicate the radius of the clouds. The
Sun, with the coordinate of ($X, Y$) = ($-$8.34, 0)\,kpc, is located at the centre of the plot. The solid red lines are the best-fitting spiral arm models of the Scutum, Sagittarius, Local, and Perseus Arms from \citet{Chen2019}.}
\label{xy}
\end{figure}

\begin{figure}
\centering
\includegraphics[width=0.45\textwidth, angle=0]{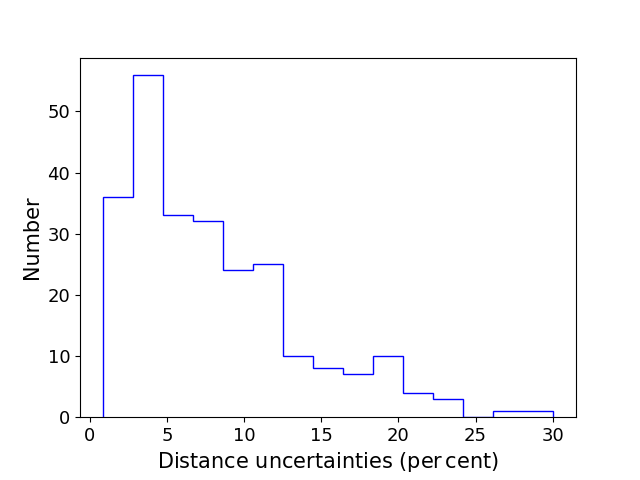}
\caption{Distribution of the distance uncertainties of our catalogued molecular clouds.}
\label{errdis}
\end{figure}

\section{Method}
\label{method}
In the current work, we have adopted a method similar as in the work of \citet{Chen2020}. The method comprises three steps: the first step is to isolate the individual molecular clouds from the 3D extinction maps; the second step to estimate the distances of the identified molecular clouds; and the third step to calculate their physical properties.

\subsection{Molecular cloud identification}
To identify the molecular clouds, we firs interpolate the extinction maps of Guo et al. to a 3D data cube of differential extinction values contributed by the local dust grains, $\Delta A_r$ (in units of mag\,kpc$^{-1}$), of which the three dimensions correspond to the Galactic coordinates $l$ and $b$, and the distance $d$. We adopt resolutions of 0.2\degr, \ 0.1\degr \ and 0.2\,kpc in $l$, $b$ and $d$, respectively. The nearest neighbour inter-
polation is adopted in the current work. The {\sc python} program {\sc dendrogram} \citep{Rosolowsky2008} is applied to the 3D data cube to identify the molecular clouds. Dendrograms are topological representations of the hierarchical structure of nested isosurfaces in $N$-dimensional intensity data. The dendrogram algorithm has three input parameters: (i) ``min\_value'', a minimum absolute intensity threshold required to mask any structures that peak below it; (ii) ``min\_delta'', a minimum intensity contrast required to exclude any local maxima identified because of the noise; and (iii) ``min\_npix'', a minimum number of pixels (volume elements in the $lbd$ space) that a structure should contain. We have tried different sets of input parameters. We have tried different sets of input parameters and visually check the resulting clumps. Eventually, we set min\_value = 0.06\,mag\,kpc$^{-1}$, min\_delta = 0.08\,mag\,kpc$^{-1}$ and min\_npix = 20. Independent structure `leaves' produced by the dendrogram algorithm represent regions of density enhancements and are identified as isolated molecular clouds in the current work.
We note that in the current work, we have applied the dendrogram
algorithm in the $l, b$ and $d$ space. As the pixel volumes increase
with the distances, our algorithm would be less sensitive to the
small distant clouds.

\subsection{Distance measurement}
Since the distance resolution of the adopted extinction map is only
0.2\,kpc, the cloud distances yielded by the dendrogram algorithm
have relatively large errors. In the current work, we estimate the cloud distances from the extinction profiles of the sightlines toward the individual clouds \citep[e.g.][]{Schlafly2014, Chen2017, Yu2019, Chen2020, Chen2020b, Guo2021b}. The distance to a cloud can be estimated by locating the position where there is a sharp increase (`jump') of extinction in the extinction profile, caused by the high dust density of the molecular cloud.

The $lbd$ 3D spatial ranges of the individual clouds are provided by the dendrogram algorithm. We select all stars in the catalogue of \citet{Guo2021a} that fall within the $l$ and $b$ ranges of a given cloud. We then fit the extinction and distance relation $A_r(d$) of the sightline toward the cloud, established by the distances and extinction values of the selected stars, to estimate the distance of the cloud. In the current work, we assume that the increase of the extinction within the distance range of the cloud is all caused by the dust grains in the cloud \citep{Chen2020, Guo2021b}. The extinction profile within the distance range of the cloud is then fitted by,
\begin{equation}
A_r(d) = A_r^0 + A_r^1(d) \quad $if$\quad d_{\rm min} < d < d_{\rm max},
\label{err1}
\end{equation}
where $A_r^0$ is the foreground extinction, $A_r^1(d)$ is the extinction contributed by the dust in the cloud at distance $d$, and $d_{\rm min}$ and $d_{\rm max}$ are respectively the lower and upper limits of the distance range of the cloud provided by the dendrogram algorithm. Assuming a simple Gaussian distribution of dust in the cloud \citep{Chen2017, Zhao2020, Guo2021b}, we have,
\begin{equation}
A_r^1(d) = \frac{\delta A_r}{2}[1 + {\rm{erf}}(\frac{d-d_0}{\sqrt{2}\delta d})],
\label{err2}
\end{equation}
where $\delta A_r$ is the total $r$-band extinction contributed by the dust in the cloud, $d_{\rm 0}$ the distance of the cloud and $\delta d$ the width of the
extinction jump.

For each cloud, the foreground extinction $A_r^0$, distance of the cloud $d_{\rm 0}$, the `jump' width $\delta d$ and amplitude $\delta A_r$ are free parameters to fit. As there could be more than one cloud along a given line of sight, only the extinction profile within the distance range $d_{\rm min} -$ 0.2\,kpc $< d < d_{\rm max}$ + 0.2\,kpc is fitted to avoid the contamination of other clouds. We employ a sliding window of width 25\,pc with a step of 50\,pc to obtain median values of extinction for the individual distance bins. The binned median extinction values are then fitted using the extinction model described above with the {\sc python} Markov Chain Monte Carlo (MCMC) algorithm {\sc emcee} \citep{Foreman2013}. We adopt the likelihood
\begin{equation}
\mathcal{L} = \prod_{n=1}^{N} \frac{1}{\sqrt{2\pi}\sigma_{n}} {\rm{exp}}(\frac{-({A_{\rm{r}}}^n-{A_{\rm{r}}}^n(d))^2}{2{\sigma_n}^2}) ,
\label{like}
\end{equation}
where ${A_{\rm{r}}}^n$ and ${A_{\rm{r}}}^n(d)$ are the $n$th binned median $r$-band extinction values derived from the individual stars and that modeled by Equation \eqref{err1}, respectively, and $\sigma_n$ is the standard error of the $n$th binned median extinction. We adopt the priors
\begin{multline*}
P(A_r^0, d_{\rm 0}, \delta A_r, \delta d) =   \begin{cases} 1 \quad \text{if} \,\,\begin{cases} 0.0 \leq A_r^0<A_{r,\rm {max}}, \\
d_{\rm min}-0.2\,{\rm kpc}<d_{\rm 0}<d_{\rm {max}}+0.2\,{\rm kpc}, \\
0.0<\delta A_r<A_{r,\rm {max}}, \\
0.0 <\delta d< d_{\rm max}-d_{\rm min},   \\
 \end{cases}\\
 0 \quad \text{else}.
 \end{cases}
\end{multline*}
$A_{r,\rm {max}}$ is the maximum extinction value of all the selected stars.

The Monte Carlo method \citep{Wall2003} is adopted to estimate the uncertainties of the resultant parameters. For each cloud, we randomly generate 300 samples of stars sampling the distance and extinction errors. The fit algorithm is applied to all the samples and the rms scatters of the best-fit parameters are adopted as their errors.

We have also adopted other bin sizes, such as\,30 pc, 50\,pc and 70\,pc, for the fitting of the extinction profiles, which yield very similar results. We note that in the current work, we treat the molecular cloud as a whole and fit the median extinction and distance profile of the cloud, instead of extinction profiles of the individual stars \citep{Schlafly2014, Zucker2019, Chen2020}. For comparison, we have selected four example clouds and estimated their parameters using both our method and the method from \citet{Chen2020}. Results from the two
approaches are almost identical.
\begin{figure}
\centering
\includegraphics[width=0.45\textwidth, angle=0]{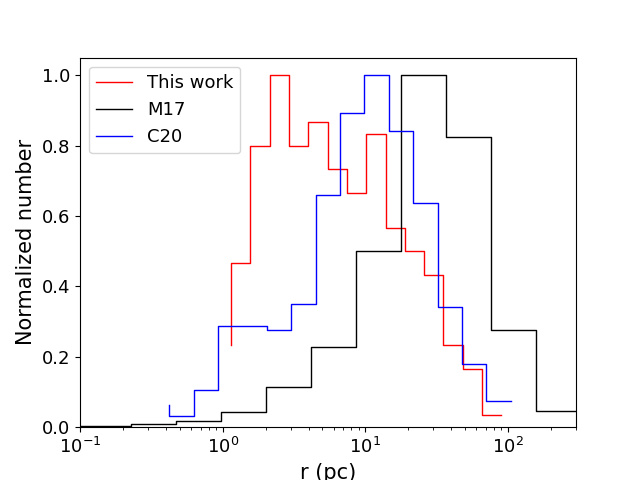}
\includegraphics[width=0.45\textwidth, angle=0]{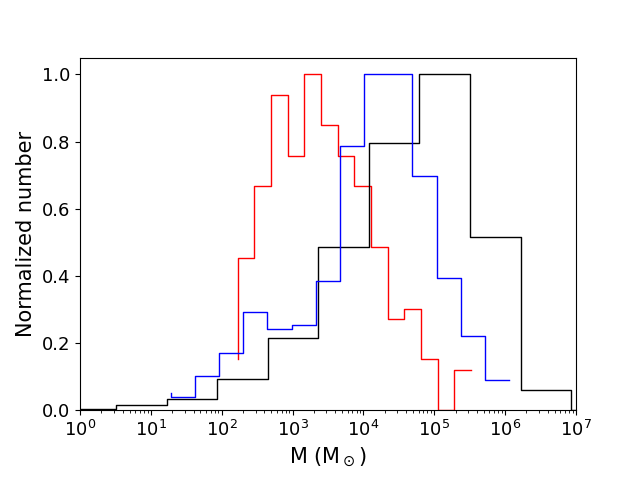}
\includegraphics[width=0.45\textwidth, angle=0]{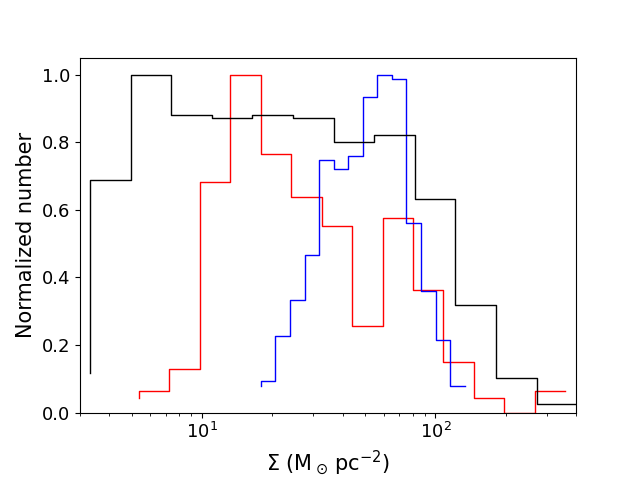}

\caption{Normalized histograms distributions of the physical properties, including radii (upper panel), masses (middle panel), and surface mass densities (bottom panel), of our catalogued molecular clouds (red), and of those from \citet[][ black]{Miville2017} and from \citet[][ blue]{Chen2020}.}
\label{hist}
\end{figure}

\begin{figure}
\centering
\includegraphics[width=0.45\textwidth, angle=0]{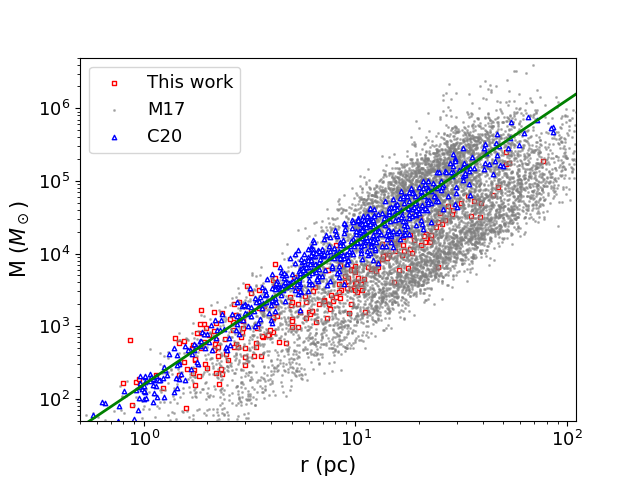}
\caption{Radius-mass relation for clouds cataloged in this work (red squares) and for those from \citet[ gray dots]{Miville2017}, and from \citet[ blue triangles]{Chen2020}. The green line is the best-fit radius-mass relation for clouds presented in the current work and in \citet{Chen2020}.}
\label{M_r}
\end{figure}

\begin{figure}
\centering
\includegraphics[width=0.48\textwidth, angle=0]{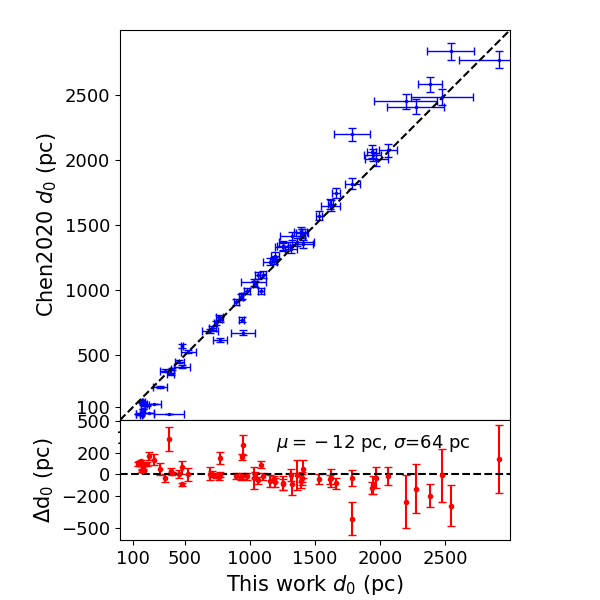}
\caption{Comparison of our cloud distances with those from \citet{Chen2020}. A dashed line denoting complete equality is over-plotted to guide the eyes.}
\label{comparedis}
\end{figure}

\begin{figure}
\centering
\includegraphics[width=0.45\textwidth, angle=0]{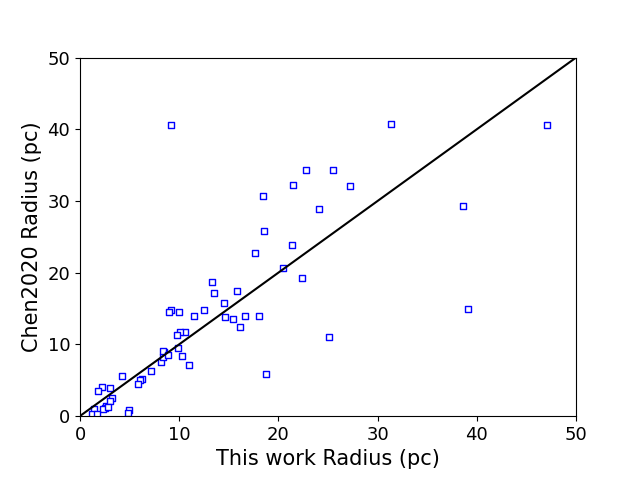}
\includegraphics[width=0.45\textwidth, angle=0]{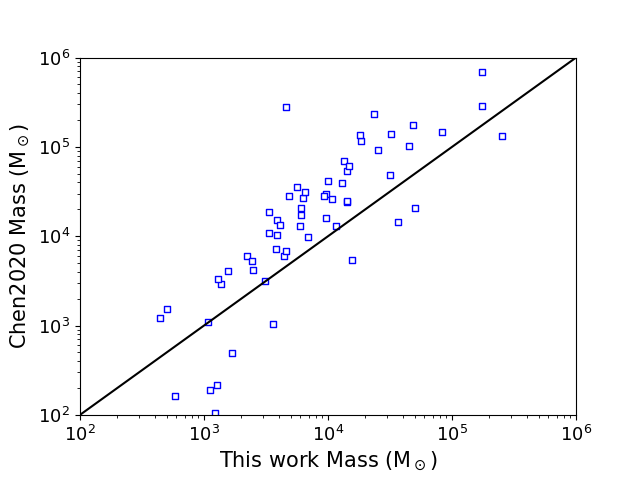}
\caption{Comparison of physical parameters, including radius (upper panel) and mass (bottom panel) obtained in the current work with those of \citet{Chen2020}. A black lines denoting complete equality is over- plotted to guide the eyes  in each panel.}
\label{comparepara}
\end{figure}

\subsection{Physical parameters}
\newcommand{\RNum}[1]{\uppercase\expandafter{\romannumeral #1\relax}}
With the distances derived, we are then able to compute the properties of the identified molecular clouds, following \citet{Chen2020}.

(i) Solid angle $\Omega$, in units of deg$^2$,
\begin{equation}
\Omega = \sum_{i=0}^{N}\Delta l \Delta b \,{\rm cos} b_i, 
\end{equation}
where $\Delta l= $ 0.2\degr\ and $\Delta b=$ 0.1\degr\ are respectively the $l$ and $b$ angular resolution of our 3D data cube, $i$ is the pixel index and $b_i$ the Galactic
latitude of the $i$th pixel.

(ii) Area S, in units of pc$^2$,
\begin{equation}
S = \Omega\,d_0^2. 
\end{equation}

(iii) Linear radius $r$, in units of pc,
\begin{equation}
r= \sqrt{\frac{S}{\pi}}.
\end{equation}

(iv) Mass $M$, in units of $M_\odot$,
\begin{equation}
M = \frac{\mu m_{_{\rm H}}A_VS}{DGR}.
\end{equation}

(v) Surface density $\Sigma$, in units of $M_\odot$\,pc$^2$,
\begin{equation}
\Sigma = \frac{M}{S}. 
\end{equation}

In the current work, we adopt $\mu$ = 1.37 \citep{Lombardi2011} is the mean molecular weight, $m_{_{\rm H}}$ the mass of hydrogen atom, $A_V$ the optical extinction of the cloud converted from $A_r$ using the extinction law of \citet{Yuan2013}, $A_V$ = 1.15$A_r$, and $DGR$ = 4.15$\times$ 10$^{-22}$\,mag\,cm$^2$ \citep{Chen2015} the dust-to-gas ratio. \citet{Chen2015} have combined the extinction
map with the H{\sc i} and CO observations to study the correlation between the dust extinction and interstellar gas content at inter-mediate and high Galactic latitudes. The average value of $DGR$ deduced by \citet{Chen2015}, 4.15$\times$ 10$^{-22}$\,mag\,cm$^2$, is slightly smaller than the most cited value derived by \citet{Bohlin1978}, 5.3$\times$ 10$^{-22}$\,mag\,cm$^2$. Literature values of $DGR$ vary by a factor of 2, from $\sim$ 3.5 to 7.0$\times$ 10$^{-22}$\,mag\,cm$^2$ \citep{Shull1985, Diplas1994, Liszt2014, Zhu2017}. In addition,
\citet{Chen2015} found a factor of $\sim$ 2 cloud-to-cloud variations
in the $DGR$ values amongst the individual molecular clouds.

\section{Result}
In the current work, we have identified 250 molecular clouds in the southern sky from the Guo et al. 3D extinction maps. Distances as well as other physical properties, such as the radius, mass and surface density, of the individual clouds are derived. The spatial distribution map of these clouds is presented as Fig.\,\ref{contour}. A large fraction of the identified isolated clouds locate near the Galactic plane. However, due to the missing data of SMSS DR1 in a few regions close to the Galactic plane, we note that for some clouds, only parts of them are identified by our algorithm. Radii, masses and surface densities of those clouds may be biased due to the incompleteness of data. However, their distances estimates should not be affected. 71 out of the 250 identified clouds locate at high Galactic latitudes ($b < -10\degr$). Most of these clouds are fully covered in our work. The full result is listed in Table\,\ref{catalogue}. Each row of Table\,\ref{catalogue} contains the information of an identified molecular cloud: position $l$ and $b$, solid angle $\Omega$, radius $r$, distance $d_0$, width of the extinction jump $\delta d$, total extinction $\delta A_r$, mass $M$ and surface density $\Sigma$.

\subsection{Distances of the molecular clouds}
Our algorithm of distance determination was successfully applied to all the identified 250 clouds. Typically, we are able to select $\sim$ 2,500 stars in the lines of sight overlapping with a given molecular cloud from the catalogue of \citet{Guo2021a}. For all the clouds, significant extinction jumps produced by dust in the corresponding molecular clouds are clearly visible in the distance ranges deduced by the dendrogram algorithm and they are nicely fitted by our extinction model. We show the distance fitting results of four example molecular clouds in Fig.\,\ref{distance} and Fig.\,\ref{twojump}. Analogous figures for all the catalogued molecular clouds are available online\footnote{\url{http://paperdata.china-vo.org/guo/cloud/cloud_250.pdf}}.

In the upper row of Fig.\,\ref{distance}, we show the distance determination of a newly identified molecular cloud that locates at high Galactic latitudes in the southern sky (Cloud No.\,4; $b \sim -36$\,\degr). The distance, $d =$ 143 $\pm$ 7\,pc, shows it is very close to the Sun, as expected. The number of stars that are foreground to the cloud is relatively small due to the saturation effects of the photometric data. But as shown in the Figure, we have enough foreground stars to identify the position of the extinction jump with  a relatively small uncertainty. The width of the extinction jump is only $\delta d =$ 41\,pc. It is mainly contributed by the intrinsic depth of the cloud, as the distance uncertainties of those nearby stars given by  Gaia DR2 are small. In the bottom row of Fig.\,\ref{distance}, we show the distance determination of the molecular cloud Mon\,R2 (No.\,180) located at $b \sim -$9\degr.\ The resultant distance of the cloud is $d_0 =$ 928 $\pm$ 16\,pc. The width of the extinction jump is $\delta d$ = 184\,pc, larger than that of the closer cloud No.\,4. This is mainly due to the larger distance uncertainties for more distant stars  from the Sun.

In Fig.\,\ref{twojump}, we show the results of our extinction profile fitting procedure for two molecular clouds located close to the Galactic plane: Clouds No.\,182 and 237, both at $b \sim -$5\degr.\ Both clouds are not fully covered by the 3D extinction maps of Guo et al. We are only able to identify parts of them. Along a given line of sight close to the Galactic plane, there are usually more than one molecular clouds that locate at different distances. This leads to more than one extinction jump in the extinction profile. For the two example clouds shown in Fig.\,\ref{twojump}, two significant extinction jumps are visible in each profile. Thanks to the distance limits provided by the {\sc dendrogram} algorithm, we are able to isolate the true extinction jumps of the catalogued clouds and estimate their accurate distances. The resultant distances of Cloud No.\,182 and 237 are, respectively, $d_0 =$ 781$\pm$ 43\,pc and 1,957 $\pm$ 61\,pc, and the widths of the extinction jumps are $\delta d =$ 103 and 302\,pc, respectively.

We plot the centre longitudes $l$ and latitudes $b$ of all the catalogued molecular clouds against distances $d_0$ in Fig.\,\ref{lb_dis}. The distances to our identified clouds range between $d_0 \approx$ 70\,pc and 3,500\,pc. The distribution of our catalogued clouds shows clearly the large-scale structure of the Galactic disk (the Galactic spiral structure; see also Sect.\,5.2 of \citealt{Chen2020}). High Galactic latitude ($|b| >$10\degr) molecular clouds are mostly local ($d_0 <$ 200\,pc). For regions of Galactic longitudes 0\degr $< l < $50\degr\ , due to the lack of data at low Galactic latitudes ($|b|<$10\degr), our identified molecular clouds are all local. The widths of the extinction jumps $\delta$d increase with the distances of the clouds. They are the results of the combination of the intrinsic depths of the clouds and the distance uncertainties
of the individual stars. Assuming that the molecular clouds are spherical, the depths of the clouds are usually less than 100\,pc. Thus for distant clouds, the widths of the extinction jumps $\delta$d are mainly dominated by the distance uncertainties of the individual stars.

Fig.\,\ref{xy} plots the spatial distribution of our catalogued molecular clouds in the $X-Y$ plane. All our sample molecular clouds
are located in the third and fourth Galactic quadrants. The distribution of molecular clouds from \citet{Chen2020} is also overplotted in the Figure for comparison. \citet{Chen2020} have isolated 567 molecular clouds from the 3D extinction maps of \citet{Chen2019} and derived their distances and physical
properties. The molecular clouds in the southern sky identified in the current work have a very similar spatial distribution with those from \citet{Chen2020}. They may be spatially associated with the spiral arms and the arm branches or spurs of the Milky Way.

Owing to the high precision of the Gaia DR2 parallax measurements for stars within 3.5\,kpc from the Sun and the large number of stars that trace the extinction profiles of sightlines toward the individual clouds, we are able to obtain distances of the catalogued clouds with high precision. In Fig.\,\ref{errdis}, we show a histogram distribution of the resultant distance uncertainties of the individual clouds. The median distance uncertainty of the catalogued clouds is $\sim$ 7\,per\,cent. About 70 per\,cent of the catalogued clouds have distance errors smaller than 10 per\,cent. A few clouds have large relative distance uncertainties ($>$20 per\,cent). They are mainly local molecular clouds with very small distances. For these clouds, the quality of extinction profile fits is relatively poor due to the relatively small number of foreground stars.

\subsection{Physical properties of the molecular clouds}
The derived physical properties, including radius $r$, mass $M$, and surface density $\Sigma$, of the individual clouds are also listed in Table\,\ref{catalogue}. For the catalogued clouds, their radii range from $\sim$ 1 to 77\,pc, with a median value of 5\,pc; their masses range from $\sim$ 7.5$\times 10^1$ to 2.5$\times 10^5$\,M$_\odot$, with a median value of 1.7$\times 10^3$\,M$_\odot$; and their surface densities from $\sim$ 3 to 310\,M$_\odot$\,pc$^{-2}$, with a median value 20\,M$_\odot$\,pc$^{-2}$. The distributions of the radii, masses and surface densities of the molecular clouds are presented in Fig.\,\ref{hist}. For comparison, we also show the distributions of the corresponding properties of clouds from \citet{Miville2017} and \citet{Chen2020}. \citet{Miville2017} identified 8,107 molecular clouds and assigned their properties based on the CO data of \citet{Dame2001}. \citet{Chen2020} isolated 567 molecular clouds in the Galactic disk and derived their properties based on the 3D extinction maps and stellar catalogue of \citet{Chen2019}. Our catalogued clouds have property  ranges similar to those of the previous studies. However, our catalogued clouds have smaller peak radius and mass than those from \citet{Miville2017} and \citet{Chen2020}. This is largely due to the incomplete sky coverage of the molecular clouds close to the Galactic plane.

In Fig.\,\ref{M_r}, we plot the masses of our catalogued clouds against their radii. The mass-size relations derived from \citet{Miville2017} and \citet{Chen2020} are also over-plotted in the Figure for comparison. The mass-size relation of our catalogued molecular clouds of radii smaller than $\sim$ 10\,pc or masses smaller than $\sim$ 5$\times 10^3$\,M$_\odot$ is consistent with those from the previous studies. While for the molecular clouds of large masses/radii in our catalogue, the mass-size relation is biased to smaller masses compared to the literature. Since most of those large mass clouds are located close to the Galactic plane, they are not fully covered by the  extinction maps of \citet{Guo2021a} due to the lack of good-quality SMSS DR1 data \citep{Wolf2018}. The missing parts of the clouds are usually those closer to the Galactic plane, where the extinction is high. This indicates that we are only able to isolate the low density parts of those Galactic plane clouds, leading to a bias in the  resultant mass-size relation.

\section{Discussion}

\begin{table*}
  \centering
  \caption{Comparison of our cloud distances with the literature values.}
  \begin{tabular}{rrrrrr} 
    \hline
    \hline
    Name        & ID  & $l$       & $b$       & $d_0$            & Literature $d_0$                                     \\
                &     & ($\degr$) & ($\degr$) & (pc)             & (pc)                                                 \\
    \hline
    
Ophiuchus & 115 & 358.09 & 15.54 & 139 $\pm$ 10 & 139$^a$, 125$^b$, 142$^c$, 141$^d$\\
Ophiuchus & 116 & 350.12 & 16.83 & 148 $\pm$ 4 & 139$^a$, 125$^b$, 142$^c$, 141$^d$\\
Ophiuchus & 120 & 355.81 & 15.98 & 134 $\pm$ 27 & 139$^a$, 125$^b$, 142$^c$, 141$^d$\\
Ophiuchus & 125 & 356.22 & 17.15 & 134 $\pm$ 18 & 139$^a$, 125$^b$, 142$^c$, 141$^d$\\
Orion & 141 & 206.65 & $-$19.33 & 418 $\pm$ 3 & 484$^b$, 389$^e$, 414$^f$, 371$^g$\\
Orion & 142 & 212.92 & $-$19.38 & 423 $\pm$ 27 & 484$^b$, 389$^e$, 414$^f$, 371$^g$\\
Orion & 143 & 208.42 & $-$16.59 & 444 $\pm$ 8 & 484$^b$, 389$^e$, 414$^f$, 371$^g$\\
Orion & 144 & 207.06 & $-$14.98 & 429 $\pm$ 5 & 484$^b$, 389$^e$, 414$^f$, 371$^g$\\
Orion & 173 & 212.93 & $-$18.56 & 486 $\pm$ 48 & 484$^b$, 389$^e$, 414$^f$, 371$^g$\\
Orion & 174 & 210.48 & $-$18.22 & 438 $\pm$ 28 & 484$^b$, 389$^e$, 414$^f$, 371$^g$\\
Orion & 175 & 210.06 & $-$17.37 & 459 $\pm$ 30 & 484$^b$, 389$^e$, 414$^f$, 371$^g$\\
Orion & 176 & 211.19 & $-$14.17 & 483 $\pm$ 58 & 484$^b$, 389$^e$, 414$^f$, 371$^g$\\
Chamaeleon & 26 & 304.53 & $-$21.94 & 193 $\pm$ 12 & 192$^h$, 184$^i$, 188$^i$\\
Chamaeleon & 32 & 294.25 & $-$19.32 & 163 $\pm$ 30 & 192$^h$, 184$^i$, 188$^i$\\
Chamaeleon & 33 & 304.42 & $-$19.07 & 209 $\pm$ 9 & 192$^h$, 184$^i$, 188$^i$\\
Chamaeleon & 34 & 299.98 & $-$17.24 & 161 $\pm$ 11 & 192$^h$, 184$^i$, 188$^i$\\
Chamaeleon & 35 & 302.45 & $-$17.56 & 197 $\pm$ 19 & 192$^h$, 184$^i$, 188$^i$\\
Chamaeleon & 38 & 294.29 & $-$16.9 & 195 $\pm$ 8 & 192$^h$, 184$^i$, 188$^i$\\
Chamaeleon & 39 & 296.8 & $-$15.98 & 186 $\pm$ 4 & 192$^h$, 184$^i$, 188$^i$\\
Chamaeleon & 40 & 302.94 & $-$16.4 & 192 $\pm$ 27 & 192$^h$, 184$^i$, 188$^i$\\
Chamaeleon & 44 & 302.82 & $-$14.27 & 175 $\pm$ 5 & 192$^h$, 184$^i$, 188$^i$\\
Chamaeleon & 45 & 294.24 & $-$14.7 & 175 $\pm$ 9 & 192$^h$, 184$^i$, 188$^i$\\
Chamaeleon & 48 & 298.9 & $-$13.47 & 177 $\pm$ 6 & 192$^h$, 184$^i$, 188$^i$\\
Lupus & 104 & 353.62 & 14.22 & 153 $\pm$ 9 & 165$^j$, 140$^k$, 155$^l$, 160$^m$, 175$^n$\\
Lupus & 107 & 349.41 & 14.66 & 173 $\pm$ 18 & 165$^j$, 140$^k$, 155$^l$, 160$^m$, 175$^n$\\
Lupus & 110 & 350.77 & 14.5 & 149 $\pm$ 31 & 165$^j$, 140$^k$, 155$^l$, 160$^m$, 175$^n$\\
Lupus & 116 & 350.12 & 16.83 & 148 $\pm$ 4 & 1165$^j$, 140$^k$, 155$^l$, 160$^m$, 175$^n$\\
Lupus & 120 & 355.81 & 15.98 & 134 $\pm$ 27 & 165$^j$, 140$^k$, 155$^l$, 160$^m$, 175$^n$\\
Mon\,R2 & 180 & 218.99 & $-$8.87 & 928 $\pm$ 15 & 952$^b$, 905$^g$, 830$^o$,  867$^p$\\
Mon\,R2 & 194 & 219.58 & $-$10.73 & 875 $\pm$ 30 & 952$^b$, 905$^g$, 830$^o$,  867$^p$\\

    \hline
  \end{tabular}
    \parbox{\textwidth}{\footnotesize \baselineskip 3.8mm
    {\it Note.} References: $^a$ \citet{Mamajek2008}, $^b$\citet{Schlafly2014}, $^c$\citet{Ortiz2017}, $^d$\citet{Ortiz2018}, $^e$\citet{Menten2007}, $^f$\citet{Sandstrom2007}, $^g$\citet{Lombardi2011}, $^h$\citet{Roccatagliata2018}, $^i$\citet{Voirin2018}, $^j$\citet{Franco1990}, $^k$\citet{Hughes1993}, $^l$\citet{Lombardi2008}, $^m$\citet{Dzib2018}, $^n$\citet{Galli2020} $^o$\citet{Herbst1976}, $^p$\citet{Chen2020}.
  }
  \label{famousclouds}
\end{table*}

\subsection{Comparison with \citet{Chen2020}}
Here we compare our derived distances and physical properties of the molecular clouds with those obtained by \citet{Chen2020}. There are 65 molecular clouds that are in common to the two studies. In Fig.\,\ref{comparedis}, we compare our derived distances with those of \citet{Chen2020}. The agreement is very good. The mean and scatter of the differences are only $-$12\,pc and 64\,pc, respectively.

In Fig.\,\ref{comparepara} we compare our resultant radii and masses of clouds with those of \citet{Chen2020}. Our radius and mass estimates,
correlate well with those of Chen et al. but are systematically lower. This is again largely due to the incomplete sky coverage of the molecular clouds in the current work.

\subsection{Distances to some well studied giant molecular clouds}
We have selected 5 famous and well studied giant molecular
clouds from our catalogue. The distance estimates of those 5 clouds
in the literature are collected and compared to our values for verify our results. The comparison is presented in Table\,\ref{famousclouds}. Overall, our results are well consistent with the values in the literature.

The Ophiuchus molecular cloud complex is one of the most nearby giant
molecular clouds. In this work, we have isolated 4 clouds that should belong
to the complex, which has a distances ranging from 134 and 148\,pc. In
the earlier studies, the distance of the Ophiuchus cloud was estimated
to be $d_0 =$ 139 $\pm$ 6\,pc \citep{Mamajek2008}, 125\,pc \citep{Schlafly2014}, 142\,pc \citep{Ortiz2017} and 141\,pc \citet{Ortiz2018}, which are all in good agreement with our results.

The Orion molecular cloud complex is one of the most extensively studied molecular clouds in our Milky Way, covering a large sky area. We have identified 8 isolated clouds belonging to this complex. They have distances between $\sim$ 420 and\,480 pc, in good agreement with the estimates of \citet[][ 389 $\pm$ 23\,pc]{Menten2007}, \citet[][ 414 $\pm$ 7\,pc]{Sandstrom2007} and \citet[][ $\sim$ 360 to 630\,pc]{Schlafly2014}, but slightly larger than the result of \citet[][ 371 $\pm$ 10\,pc]{Lombardi2011}.

Chamaeleon is a well studied giant cloud south of the Galactic plane ($l \sim$ 300\degr,\ $b \sim -$15\degr). We have identified 11 clouds probably belonging to it in the current work. They have distances around $\sim$ 182\,pc, in very good agreement with the results in the literature: 182\,pc \citep{Roccatagliata2018} and 188 and 184\,pc \citep{Voirin2018}.

The Lupus molecular cloud complex is one of the largest low-mass star-forming regions \citep{Comeron2008}. Previous distances estimate of the Lupus complex are 165 $\pm$ 15\,pc \citep{Franco1990}, 140 $\pm$ 20\,pc
\citep{Hughes1993}, 155 $\pm$ 8\,pc \citep{Lombardi2008}, $\sim$ 160\,pc \citep{Dzib2018} and 175 $\pm$ 1\,pc \citep{Galli2020}. Those values compare very well with our determination ($\sim$ 135 to 175\,pc).

The Monoceros R2 region (Mon\,R2) is a chain of reflection nebulae extending over 2\degr \ in the constellation Monoceros \citep{Van1966, Lombardi2011}. \citet{Herbst1976} estimated a distance of 830 $\pm$ 50\,pc to the Mon\,R2. \citet{Lombardi2011} refined its distance to 905 $\pm$ 37\,pc. \citet{Schlafly2014} provided distances to different parts of the cloud, ranging from $\sim$ 830 to 1040\,pc. \citet{Chen2020} obtained a distance of $\sim$ 867\,pc. All the results are well consistent with ours ($\sim$ 901\,pc).

\section{Summary}
In the current work, we have created a catalogue of 250 molecular
clouds in the southern sky identified from the 3D extinction maps of \citet{Guo2021a} by a hierarchical structure identification method. 71 of the clouds locate at high Galactic latitudes of the souther sky ($b < -$10\degr).

For each cloud, we have selected stars that locate in the $l$ and $b$
ranges of the cloud from the star catalogue of \citet{Guo2021a}
and fitted their extinction and distance relation to find the distance of the cloud. Thanks to the accurate parallax measurements of the individual stars from the Gaia DR2 and the large number of stars that trace the extinction profiles, we are able to estimate the distances to our catalogued clouds with high precision. The typical distance uncertainty is $\sim$ 7 per\,cent. We have also estimated the properties, including the area, radius, mass and surface mass density of the clouds. Our resultant distances and properties of the clouds are consistent with those from the literature. The catalogue is available in the online version of this paper and also at weblink {\url{http://paperdata.china-vo.org/guo/cloud/cloud_para.fits}}.

\section*{Acknowledgements}
This work is funded by the National Natural Science Foundation of China (NSFC) No.\,12173034 and 11833006, Yunnan University grant No.\,C176220100007, C619300A034 and 2020288, National Key R\&D Program of
China No.\,2019YFA0405500. We acknowledge the science research
grants from the China Manned Space Project with No.\,CMS-CSST-2021-A09, CMS-CSST-2021-A08 and CMS-CSST-2021-B03. This research made use of astrodendro, a Python package to compute dendrograms of Astronomical data (http://www.dendrograms.org/).

\section*{Data availability}
The data underlying this article are available in the article and in its online supplementary material.

\bibliographystyle{mnras}
\bibliography{ms}

\end{document}